Invisible Labor: The Backbone of Open Source Software

Robin A. Lange, Anna Gibson, Milo Z. Trujillo, & Brooke Foucault Welles

Abstract

Invisible labor is an intrinsic part of the modern workplace, and includes labor that is undervalued or unrecognized such as creating collaborative atmospheres. Open source software (OSS) is software that is viewable, editable and shareable by anyone with internet access. Contributors are mostly volunteers, who participate for personal edification and because they believe in the spirit of OSS rather than for employment. Volunteerism often leads to high personnel turnover, poor maintenance and inconsistent project management. This in turn, leads to a difficulty with sustainability long term. We believe that the key to sustainable management is the invisible labor that occurs behind the scenes.  It is unclear how OSS contributors think about the invisible labor they perform or how that affects OSS sustainability. We interviewed OSS contributors and asked them about their invisible labor contributions, leadership departure, membership turnover and sustainability. We found that invisible labor is responsible for good leadership, reducing contributor turnover, and creating legitimacy for the project as an organization.

# Introduction

Open source software (OSS) projects are an integral part of many collaborations, from the technical bases of Fortune 500 companies, to facilitating international research, to game modding communities. In research, OSS projects can make life-saving work possible by connecting expertise from disparate areas. For example, cancer researchers may understand all of the gene pathways in a cell, but they do not necessarily have time to learn how to code a program to analyze their results. OSS projects can fill these gaps and allow researchers to progress their work without every researcher needing to learn how to code. Unfortunately, OSS projects are frequently abandoned and development teams face difficulties with attrition and burnout (Avelino et al., 2019).

The problems with the sustainability of OSS projects are baked into the cultural and technical premise of open source itself. The OSS ecosystem was -- and in some ways still is -- billed as a technological meritocracy and utopia (Majid, 2024). OSS projects allow people from all over the world to work together to create free projects that are available to anyone with an Internet connection. People who choose to work on these projects can gain valuable experience in coding and, theoretically, anyone who can code can then add functionality to projects that will grant them recognition by people in the tech industry (Lerner & Tirole, 2000; Oreg & Nov, 2007). In addition, projects often have corresponding communities that give feedback which can then be integrated directly into the code itself. These communities are made up of users and contributors; the users can become contributors, and project maintainers may eventually invite contributors to join leadership based on their contributions. This interactivity can allow contributors to quickly respond to the users' needs and create interconnected communities.

In practice, OSS does not always live up to its lofty ideals. Despite being advertised as a meritocracy, OSS projects do not usually have a diverse contributor base (Demby, 2013; Nafus, 2012). While contributors may care about the ideals of OSS, they do not necessarily have the leadership skill set to maintain or grow their projects past the initial building period, preferring to abandon these projects for new ones instead. This lack of leadership investment also often means that once original contributors leave, their projects are not able to transition to new leaders successfully, because the group can struggle to make collective decisions, elect a new ruler or change their existing leadership structures to ones that fit the current needs of the project. Finally, because the software is generally freely distributed, these projects cannot raise revenue to sustain contributors. The lack of monetary compensation itself precludes many contributors from participating, as it invites participation from only those that can afford to work in their free time, or those being compensated externally (Riehle et al., 2014). This also means that people can step away at any time with no warning

Though all of these problems may appear different, they can all be tied to issues involving *invisible labor*, or the work that undergirds but is invisible to the formal market economy. Invisible labor, a term coined by Arlene Daniels (1987) to discuss women's contributions to the home has since been expanded to include labor in the workplace that goes unrecognized and uncompensated. This labor includes: behind the scenes administrative work, setting up websites and coordination between members all of which OSS contributors are required to do to sustain successful projects.

In this paper we examine factors that contribute to the longevity of OSS projects, particularly focusing on invisible labor as the key connecting thread amongst issues in leadership, funding, and contributors' viewpoints toward diversity. To study this phenomenon, we conduct a series of qualitative interviews (N =19) with current and past contributors of OSS projects. We focus on community-developed projects that welcome external contribution and are seeking funding. We find that invisible labor is essential to contributor and fiscal stability: undocumented social labor improves contributor retention, provides social legitimacy, and leverages an images of legitimacy to seek external funding.

## Literature Review

**What is OSS?**

OSS projects are coding projects that anyone can view and contribute to. The development of an OSS project might look like this: a person or a group of people has an idea for a program that they want to exist. They then code up the program and put it online where anyone can download and use it. Later, other programmers may decide they want to add a modification, they can code up the modification and request that it be merged with the existing code. Anyone who contributes in any capacity is known as a contributor. The code is freely available online for anyone to view and modify, and importantly, available for anyone to use. Well known OSS projects include the Linux operating system and LibreOffice. In contrast, the code for proprietary software is only viewed, created, and edited by its respective owner, and use often depends on purchasing the software. Examples of proprietary software include Norton Antivirus or Microsoft Office. OSS has gained massive popularity and grows every year (Daigle & Github Staff; 2024). It has become increasingly important for everyday life, and is integrated into many businesses today (Ghariwala, 2023). For businesses, the value of OSS is estimated to be 8.8 trillion USD (Hoffman et al., 2024).

Open source software started off as the norm in the 1950s and 60s, when most software was created and shared freely. Today, software is a mix of proprietary and OSS. The culture of OSS is now more established with participants believing in the ideals of OSS and organizing their projects around these ideals. Individually, some people want to contribute because they identify with the project's community (Bagozzi and Dholakia, 2006), because of a desire to learn (Bonaccorsi & Rossi, 2003, Lakhani &Wolf, 2005) or because they have a personal need from an OSS project (Lakhani &Wolf, 2005).

**Hope labor and venture labor**

The history of software development in the United States is intertwined with the neoliberalization of the U.S. economy. In her book examining the decisions of dot-com era workers around employment and labor in the late 1990s, Gina Neff (2012) introduced the concept of "venture labor" to articulate how these workers conceptualized their current work as being investment in their futures. She found that even though these workers were employed at companies, they explicitly expressed this work as entrepreneurial. Such labor, she argues, was a response from the changing U.S. economy in which risk was shifting from corporations onto individuals. Within the context of widespread financialization, changing valuation, and the introduction of flexible work, individuals needed to "bet on themselves" rather than the companies they were working for. Expanding on this idea, Kuehn and Corrigan (2013) examined the labor of unpaid digital workers, finding that often this work was conducted in the "hope that future employment opportunities may follow" (2013, p. 10), work that they named "hope labor."

Within the open source community, hope labor is rife. For contributors on OSS projects, the legitimacy of their work is achieved through the public record of code commits on sites like GitHub. These websites create public records of when contributors have created or permanently changed code in the software, signifying substantial labor related to coding. However, much of the work involved in making open source projects function lies beyond writing code, and is either undocumented or unacknowledged.

**Invisible Labor**

Arlene Daniels (1987) coined the term "invisible work" to describe work that goes unacknowledged as work because of its lack of societal salience as work, often because it is uncompensated, lacks broader visibility, and is highly gendered. Originally used to describe work in the private sphere such as homemaking, the term has expanded to include invisible labor in the workplace. In the workplace, where it is similarly undervalued because of its traditional links to women and people of color (mostly Black people) (Finley, 2020).

Kaplan (2022) classifies invisible labor into four categories: invisible teamwork, invisible physical care work, invisible emotional labor, and invisible administrative work. Invisible teamwork includes labor that maintains team cohesion, such as checking in with team members or creating social events. Invisible physical care work is about taking care of the office environment such as cleaning up after a party. Invisible emotional labor is labor that lifts up coworkers such as encouraging them to speak up in meetings or mentoring them. Finally, invisible administrative work includes necessary logistical tasks like scheduling meetings or secretarial tasks. In both traditional work and OSS work these types of labor are similarly obscured. At the same time, all of these forms of labor are necessary for projects to function but will not necessarily be visible to an outsider or easily viewable in a GitHub repository, which only records code contributions.

Visible work in OSS looks like coding contributions on GitHub, formal recognition on websites or formal recognition within an organization. However, maintainers often only acknowledge code contributions on websites and in public statements, and GitHub only tracks code contributions. Some labor may be visible but undervalued, like responding to people in community forums, while other labor like mentorship takes place in Discord servers and Slack workspaces, which is completely opaque to outsiders.

Like in many fields, invisible labor is an integral part of an open source project's everyday functioning. However, the expectation that work on OSS projects will be "worth it" for recognition and career success does not take into account the effort that must be spent on invisible labor. Meluso et al., (2024) found that half of the labor that goes into maintaining an OSS project is under-acknowledged. This means that OSS work may actually be much less lucrative than anticipated for both newcomers trying to break into coding and experienced coders that have a better sense of the high monetary value of their time. Moderation and community management, like many kinds of invisible work identified by Daniels (1987), require "feminine" soft social skills that may also be seen as low-value or unimportant next to the "masculine" technical skill of coding.

The prominence of invisible labor in OSS is a problem because it means that OSS is falling short of its ideals for contributors. When contributions go unrecognized, contributor's careers do not advance because outsiders are unable to see their contributions. When they do not feel like a valued member of the community, and are also not getting paid, they are only left with intrinsic motivations such as learning (Lakhani & Wolf, 2005) or adding functionality to a tool . This is bad for the individual and bad for the project. When contributors feel low investment due to their unseen labor it leads to high turnover. This in turn means that the OSS project has less consistency and fewer contributors who want to take on leadership roles.

In traditional work, this invisible labor is overlooked, as evidenced by its lack of compensation and formal recognition. In OSS, this work is not recognized for similar reasons that it is ignored in most workplaces. Traditional work and OSS websites may contain a contributor page that acknowledges a person's coding position but not their invisible labor related contributions. Both types of organization can also avoid formally recognizing invisible labor contributions within their organizations.

The reliance on invisible labor is not a surprise, as it is necessary for all organizations. Publicly, digital software and platforms particularly have tried to minimize the role of human labor, hiding or obscuring it with appeals to "AI power" while actually relying on large populations of low-paid "ghost workers" (Gray & Suri 2019). Commercial content moderators of social media platforms, for example, have historically been hidden from public narratives (Roberts, 2019). Other kinds of invisible labor are outsourced to volunteer users, such as Reddit moderators (Li et al, 2022).

But, there is reason to believe that OSS projects rely on more invisible labor than projects developed in traditional workplaces, because it is primarily developed using volunteer labor rather than formal employment. With nothing keeping contributors from dropping off projects it may be more important for leaders to invest in invisible teamwork that maintains team cohesion. Unlike traditional organizations where there may be a paid secretary or manager to manage meeting creation, this all falls on OSS contributors. So it would also be important to maintain invisible administrative work that would be paid and possibly more noticed in a traditional organization. For these reasons, it is important to think about how OSS leaders utilize and think about invisible labor and how that impacts OSS sustainability.

The problems that OSS faces are not unique to the OSS ecosystem. OSS shares many similarities with other communities that do not have direct and consistent financial incentives. Fandom communities (Winter et al., 2021), modding communities, Wikipedia (Halfaker et al., 2011), crowdsourcing (Brabham, 2010) and activism groups all struggle with issues around abandonment, a lack of leadership, funding and diversity. They also struggle to transition between different leaders and leadership structures. This means that the implications that we find related to OSS are broadly applicable to problems that all interest-based and non-traditionally funded organizations.

**Contributor Churn**

Leadership has a strong impact on the sustainability of OSS communities. A good leader can direct a project, mentor other contributors and create useful code themselves, while a bad leader will not properly take time to manage their project and won't onboard new members. The first leaders of OS projects are typically its founders, although as projects evolve, those founders often want to move on. When leaders leave they can create uncertain leadership and unclear development plans, which impacts long term survivability. for their OSS project and their survivability. In the founder's absence, new leaders need to take over. Leaders have more diverse skills than their members (Gadman, 2013)), and communication and team building skills are strong predictors of leadership (Hergueux & Kessler, 2022)), suggesting that while technical abilities are important, experience with invisible labor is a defining trait for successful open source projects. What this means is that skills relating to invisible labor are necessary for successful OSS projects.

Leaders leaving can disrupt projects immensely (Wasley, 1992; Singer et al., 2004). In OSS, one factor of how important a leader is to a project is called the truck factor. The truck factor is a measurement of the minimal number of contributors that a project needs to continue surviving, and in more morbid terms the minimum number of people that could get hit by a truck that would derail that project. Most OSS projects have a very low truck factor. Avelino et al. (2019) found that in their sample 57% of the OSS projects had a truck factor of 1 and 25% had a truck factor of 2. This means that for 57% of OSS projects, if their main contributor leaves the project, it will not be able to function without significant changes. Because OSS contributors are typically volunteers with no employment obligations to a project, they are more likely to leave suddenly without notice or a transition plan. As a contributor leaves they may take instrumental project knowledge with them that future contributors will need to learn. This in turn can affect project sustainability and productivity (Rashid et al., 2019).

There is not very much literature on why people who have been involved in an OSS project for a long time leave that project. Because the barriers to returning are usually very low, and most contributors can come back anytime they want, we consider a contributor to have left if they have no intention of coming back, regardless of whether they can or not. Some literature says that people who contribute to projects long term are more likely to stay if they advised other contributors and continued to make coding contributions (Fang & Neufeld, 2009). Some articles point to burn out as a reason that people leave OSS (Zuegel, 2019). Our study extends this knowledge with a qualitative interview-driven analysis of why long term open source contributors leave projects.

The survival of OSS projects also entails the onboarding of new members, some of whom will become the new leadership (Avelino et al., 2019). In most projects, contributors do not stay on long term, and can leave at any time. The reasons for this are multifaceted but largely social. Newcomers who receive few responses, little recognition and a lack of clear expectations lead newcomers to leaving OSS projects (Steinmacher et al., 2015). The more interactions that newcomers have, the more likely they are to stay with a project (Jensen et al., 2011). This is a problem for newcomers outside of the United States because they were less likely to be replied to than contributors within the United States (Jensen et al., 2011). It is important to note that most newcomers do not stay long on any project, only 20% of newcomers become long term contributors (Steinmacher et al., 2013). For those interested in joining a project, these factors are all very relevant. Taken together this means that skills that fall under invisible labor are crucial for OSS success.

**Research Questions**

1. **How do OSS project contributors understand invisible labor in their work?**
2. **How does reliance on invisible labor affect OSS project sustainability?**

# Methods

**Recruitment**

In this study we interviewed OSS contributors (N = 19) about their work. We recruited participants by sending out an IRB approved email to individual contributors and asked them to participate in our study.

To find these participants, we used lists of OSS projects that were funded by grants and other awards, for example the current Chan-Zuckerberg Essential Open Source Software for Science grant recipients. This is because funding is a big issue in the OSS community, and talking to unfunded projects may have led to answers that largely revolved around funding (Wodecki, 2024). By obtaining interviews from funded projects we hoped to get a variety of answers on the problems that the OSS community faces. On each of the pages of grant recipients we found the websites of their corresponding projects and selected projects with between 5 and 500 contributors. We hoped to find projects that were popular enough that they had attracted newcomers, but not so giant that they were fully self-sustaining organizations. If the websites had contact pages we emailed the contributors directly with IRB-approved language. In total we reached out to a total of 274 contributors, received 31 responses, and eventually conducted 19 interviews. Each interview from this particular population allowed for the iterative development of theory, allowing us to identify and expand on the concept of invisible labor in OSS work. We reached theoretical saturation with these 19 participants (Small, 2009).

In this research study, it was very difficult to negotiate access to qualified contributors. There were few OSS projects that fit our criteria and even fewer with willing members. Given these difficulties, 19 participants is an adequate number for our research study.

**Interviews**

Each interested contributor filled out a Qualtrics availability survey and were later contacted for an interview. We tried to recruit participants who were in leadership positions with the assumption that leaders had been with the project for a longer time and could give us more insight into the OSS project and community.

The interview protocol was written very generally to get a background of the OSS project and the roles of each contributor, with questions about invisible labor and leadership responsibilities. Questions were later refined to include more probing questions towards invisible labor.

Each interview took place remotely over Zoom and took between 45 minutes and 1 hour. All interviews were recorded and transcribed. All participants spoke English and received an Amazon gift card of $25 for their participation.

At the start of each interview, participants were read the consent form and consented to be a part of our study. They were then asked a series of questions. We asked them about their project, how they became involved, their roles and responsibilities and questions about the invisible labor they performed. Some participants had left their projects but were still being credited on their former project's websites. These participants were allowed to continue to participate in our study because they still had experience with OSS and because this could give us insight into factors that end individual participation.

Data consisted of interview transcripts and participants' illustrations of project organization. These data were loaded into the Dedoose software and coded with an open codebook. The first two authors met regularly during the coding process to discuss new codes and emergent themes. Invisible labor emerged as a salient construct during this process, and the interview protocol was iteratively updated to probe this construct.

# Findings

Like any organization, OSS requires a lot of invisible labor. One could make the argument that all open source work is invisible labor because none, or very little of it is formally compensated. However, because OSS projects are often seen as ways for people to get their foot in the door in the field of software development, some of this work is done by people expecting future compensation. This type of "venture labor" (Neff, 2016) takes the form of updating and improving the code of a particular project. For this paper, we frame all work that is not coding – that is, that can be viewed as a Github commit for an OSS project - as invisible labor, including posting on forums and coordinating, facilitating meetings, creating legitimacy for funding, onboarding new members, moderating and leadership.

**Variable awareness of invisible labor**

We found that the people who were primarily coding for OSS projects were less aware of the existence of invisible labor in their projects, and performed less invisible labor than other members. In contrast, OSS contributors who were in more senior or leadership positions were very cognizant of the invisible labor that they were performing. People who were performing invisible labor were doing so outside of their typical work hours, so every extra email that may have gone unnoticed for their paid job was much more salient as extra labor.

To tie this in with past research on invisible labor we again turn to Kaplan (2022)'s four types of invisible labor: invisible team work, invisible physical care work, invisible emotional labor and invisible administrative work. In investigating which types of work were used, we found that all four types were represented in our sample (see Table 1), with members frequently doing multiple types of invisible labor.

Table 1: Frequency of Invisible Labor

| Type of Labor | Example Quote | Number of Participants |
| --- | --- | --- |
| Invisible Team Work | "…I know we will share personal lives like milestones to other people. For example, I think the one I mentioned, the contributor I mentioned, just [had a milestone] and he shared that news with us. Something like that. I think also if several contributors meet together in a conference, they also share the pictures with us and they will share their meals, like skating with their skating in the, you know, in some rural areas they will share the pictures with us…" Participant 07 | 3,4, 7,11, (4 participants) |
| Invisible Physical Care Work | "…I would help people, you know, with their coding problems and their analysis problems." Participant 14 | 8, 10, 11, 13, 14, 17 (6 participants) |
| Invisible Emotional Labor | " Yeah, [I was] making sure that people know that it's ok to ask questions, it's ok to be wrong. It's ok to have opinions and thoughts." Participant 04 | 4, 10, 11, 12, 15,17, 18 (7 participants) |
| Invisible Administrative Work | "We taught a lot of workshops about our software and I led the organizational efforts for most of those and there is a significant amount of behind the scenes work | 8, 10, 11, 12, 13 (5 participants) |

| | that goes into things like that. Most of it is incredibly mundane, like making sure that people are able to get reimbursed like contributors to this, making sure that we've got all the travel visas in order, or like having a safety plan in place if we're meeting at a physical location like, ok what's the emergency evacuation route, it's that kind of stuff. I did a lot of that and I think it was often pretty invisible." Participant 04 | |
|---|---|---|

**Leadership**

The large amount of invisible labor required to maintain an OSS platform bleeds into and is strongly related to the problems that arise with leadership and long term sustainability of an OSS project. Creating a stable governance structure is crucial for OSS projects. Without governance and leadership there would be no direction for the contributors which would mean fewer bug fixes, and directions that would not necessarily reflect the needs of the community.

The reasons people joined an OSS project were varied but often boiled down to one of two categories. First, they might need the project in their own personal work and so they felt like they needed to maintain it to ensure its continued use. Second, they liked coding and wanted to contribute to a project that was important. After they joined though, most people we interviewed became more managerial than day to day contributors, a job they did not necessarily like. Participant 09 said two things about the leader of his project. First he said: "So like, he and [other leader] wrote a lot of the math library upfront… And they were kind of like… they both built a lot of the stuff that they became the… the gatekeepers for." He also said: "So [most senior leader] was very open with something like delegation, he would tell you he hates management. He doesn't really like making decisions." The leader of this project wanted to do the coding for this project, and did not like conducting management and leadership responsibilities. But, because other contributors were now joining the project, he couldn't just code, he had to manage the group so they could all work together in harmony. This participant did not want to conduct invisible labor but had to, to continue the project. Without the invisible labor of management, making decisions and scheduling, the project could not continue.

Without the pressure of employment, it's difficult for project leadership to assign "less flashy work." This means that the less glamorous work that needs to get done for a successful project, such as maintenance or bug fixes are less likely to get done than things that are flashier such as a new feature. Participant 10 said:

> "So, in terms of the assignment of tasks, I can suggest to my boss, hey, we should do this, this, and this. And I can talk to people at the meetings and try to get consensus on where other people think that things need to go. But, at the end of the day, I can't tell people, "You must do this," which has its own challenges. It makes it difficult for the things that

> nobody wants to do. You know, there's, you know, taking out the trash is something that, like, needs to get done. It's just as important as all of the other things that we do. I feel like it's important to be able to build a community where people want to do that."

A few people in our study mentioned how much the project needed people to just maintain the code and that this was difficult to sell to newcomers who were often interested in adding features that they, personally, needed. Participant 04 said:

> "I don't think there was ever a point where it was like, yeah, we don't have enough contributors. It was usually kind of the other way around where we had maybe too many contributors and not enough high level maintainers and reviewers."

This lack of maintenance meant that while new features were being added, there wasn't enough attention being paid to maintaining features that already existed. This issue of lack of maintenance connects to an issue with hope labor being based on visibility: maintenance work, by its nature, is not flashy. Hope labor relies on the visibility of contributions being recognized and rewarded in the future. The implementation of new features has much greater potential for potential recognition and celebration than maintenance work.

**Funding**

Despite the conversations of how important invisible labor is for the OSS projects, funding is also a critically important aspect in OSS sustainability. Funding allows projects to hire dedicated contributors to solely and consistently work on the OSS project. This can also mean things like up to date tutorial pages or maintained websites. Funding allows OSS projects a chance to be sustainable long term. With funding, teams can hire workers to do the undervalued labor like maintenance or documentation to pick up the slack from volunteer labor.

Obtaining funding is not an easy task and requires a lot of additional unpaid labor and skill sets not necessarily held by many contributors. Because of our sample, some of our participants were highly connected to academia, and they went about acquiring funding in academic ways, meaning that they applied for grants or found graduate students to maintain projects for research credits. While others who were not tied to academia approached funding as more of a business, for example hiring a fundraiser, trying to solicit donations or setting up their OSS project as a nonprofit. While most of our participants were active contributors to OSS, some had left after making significant contributions.

One thing that was consistent with any attempts to gain funding was that people needed their project to seem legitimate, which often required many forms of unpaid labor, such as having a functional and up to date website, a clear vision, and visible leadership. Some people tried to obtain funding in academic ways though this meant that they had to work with universities, which came with their own difficulties. Participant 15 said:

> "...what we do from day-to-day is basically follow up on the grants. Because there's a number of developers being paid to do some tasks, so we sort of check in, is this moving along, when is the next reporting coming up, writing those reports, and then also following up for the future, to see if there's new grants coming up, and write the base for the applications for them. And then in between that …we ask around on the forum and with the people we know, Hey, what would you like to see us focus on next? So sort of getting the input for an upcoming grant."

The grant writing process requires a lot of invisible labor from the community. Here, they asked the community what they needed, and incorporated it. While association with an academic lab grants some measure of legitimacy to a project, this does not automatically grant funding either. As most academics know, funding still requires grant applications and justifications about why the project is important. Overall, it is the invisible labor of maintaining a professional public presence, identifying funding opportunities, and applying to them that sustains open source projects.

Funding can have a huge impact on sustainability but reliance on a grant funding structure means that funding has a finite time limit and that projects need to continue applying for funding in cycles in order to remain viable. As an alternative source of funding, OS projects can try to gain corporate sponsorship. OSS projects can try to gain corporate funding. Some organizations are becoming increasingly aware of their reliance on a healthy OSS ecosystem for their success (such as Microsoft (Microsoft, 2024)) and are increasingly funding OSS projects. Participant 10 explained that their group founded a nonprofit in order to apply for corporate funding:

> " …we want to raise money that's independent of the university that we're attached to, we need to have some sort of chain of legitimacy for being the speakers for the project. And this is something that came out of some of the legal work that I did with [unrelated convention], where anytime you start dealing with outside contracts or money or whatnot, you need to be able to show legitimacy and we also established, as part of this, we established the Executive Committee."

In both cases, projects needed to perform huge amounts of labor to create legitimacy to then receive money that was only possible because one of their members already had pre-existing experience in this kind of funding process.

**Contributor Churn**

Onboarding new programmers is essential for projects to continue. In the long term, the old guard retire from projects, and they need newcomers to take over and become the next generation of leadership. In the shorter term, they often need newcomers to come in and maintain code, because a lot of people are only interested in writing new interesting features. However, while mentorship was a highly valuable skill for OSS leaders to have, they often didn't have it. This was another skillset under invisible labor that no one in our sample was formally taught. Participant 06 said:

> "And we weren't very good at communicating how to make contributions so people would go and build some big package or library to add to [our project], and we're like, no, no, no, no, no, it doesn't meet our coding standards."

Participant 04 said:

> "What this tended to look like, especially for let's say junior engineers that we're bringing on, we would usually allocate, or at least I would allocate, a lot of hands on time over their first week, month and six months, usually with a little bit of taper, but a lot of pair programming, a lot of hands on work to try to understand their approach, try to understand their pre-existing knowledge, where there might be holes or gaps."

Without mentorship, contributors are wasting a lot of their time trying to work on projects, but their contributions are thrown away because they do not match maintainers' standards of quality. With more mentorship in place, this could have been caught earlier, wasting less of their time if they chose to drop their contribution, or they could have been led in a direction that allowed them to make an acceptable change. The leaders here recognized they were receiving sub-par code submissions, but because they didn't have mentorship skills they failed to actually fix this problem. Mentorship here is another skill that falls under invisible labor because people who are interested have to teach it to themselves, and once they do have this skill, utilizing it often goes formally unrecognized. Participant 18:

> "...I signed up as a mentor, and just went really slowly because I didn't know much. So I would like stare at someone's question, or someone's code, and I would have to like do research. I would just sit there and figure out, and look at as many solutions as I could find until I had something valuable to say. And you know, just starting with the most basic exercises at first."

Participant 18 was not given any formal mentorship in how to mentor, and as a result it took him a long time to give feedback and they felt only qualified to help with the most basic of questions. Participant 18, shortly after this became a maintainer for this project. They clearly had the requisite skills as a coder, and if they had received more help with mentoring, they could have utilized their time more effectively.

In addition, the mentorship could have helped people contribute when they didn't yet have sufficient technical skills. Participant 04 said:

> "The places where we really ran into limitations on resources by the time I left, were more so around educational resources, which are our contributors, but it was like the kind of thing that we hadn't really broken down yet. Like, what kind of roles could exist to allow for people to contribute more tutorials or more documentation or more translations, but that was definitely a little bit of a roadblock towards the end of my time there."

This highlights that there are many ways of contributing to OSS with little experience. However, because mentorship is often lacking, potential contributors don't even know that completing tasks like creating tutorials are needed for many projects.

**Academic Open Source**

The academic nature of most of our sample also affected the invisible labor in their respective OSS projects. Academic OSS in general tends to focus on creating tools for academic researchers to use for specialized data, problems, or visualizations. Well known examples include: SciPy, Bokeh, NetworkX, PsychoPy, and Bioconda. People who participate in academic open source are often creating tools that they need for their own research, which are often specialized projects that have no other existing tools available. Participant 10 said:

"And I think this is a challenge with academic open source software, is that, for a lot of people, this is just… it's a tool that they use to do their research, and that is problematic, because then it means that a lot of sort of the infrastructure tasks – continuous integration, deployment, testing – all of that stuff that doesn't fall directly into the purview of what somebody needs to do their PhD, gets left to the wayside."

Participant 17 said: "I'm still mostly a user. My main interest is to have a good tool. If the tool was good I would never have participated."

When people are only part of a project for a specific purpose they only focus on the parts of the project they need. This might be fine for academic projects with a narrow scope, however as projects grow, the amount of infrastructure that is needed and related to invisible labor also grows. When these projects aren't able to recruit new members beyond the initial academic team, they can fall apart and lose their utility.

# Discussion

Modern infrastructure, and especially modern research, increasingly relies on OSS contributions to function. However, OSS projects have many problems that decrease their ability to function long term. Beyond the simple story of needing contributors, OSS projects frequently lack funding that allows for hiring necessary staff, lack contributors that want to take on active leadership roles, lack the ability to create legitimacy, and lack the ability to plan effectively.

Invisible labor is integral to contributor recruitment and retention, pursuing funding, and long term project planning of OSS projects. Contributors used every type of invisible labor: invisible team work, invisible physical care work, invisible emotional work, and invisible administrative work. The lack of payment or public recognition for these actions made them very salient for contributors. Leaders needed undervalued social and organizational skills to succeed that they were never formally taught, so they often had to bring in leadership experience from elsewhere. Invisible labor was integral to connecting people, keeping them involved, maintaining good coding practices behind the scenes, getting funding, mentoring future leaders and contributors, scheduling meetings, and generally maintaining the project.

**Contributor Churn**

Contributor churn is a cycle where in OSS projects can attract contributors that want to add functionality to the project, but because of a subpar structure contributors can become frustrated and leave. Subpar structure can include a lack of guidelines that lead new contributors to how to make useful contributions, unclear policies and direction and a lack of mentorship. This impacts both leadership and non-leader contributors. Leaders who do not adequately spend time on the aforementioned structure can then be burdened with reviewing unsuitable coding contributions leading to their own burnout and turning away new contributors.

Contributors to OSS are motivated by hope of future employment (hope labor), personal growth, community participation, and to improve the tools they need for work outside of OSS. The participants in this study were motivated to become leaders by all of these criteria and because of their beliefs in the OSS ethos. However, this did not always translate to staying as leaders in OSS. Some people left because their services were no longer required, they had funding but lost it, they needed to utilize their time towards more lucrative endeavors or they were focusing on other OSS projects.

Consistent with prior literature, the people who were the best at coding did not necessarily become the leaders. We found that people who are leaders did not necessarily want to be leaders, and that there were many different forms of successful leadership structures. There were different forms of successful leadership structures, from top-down benevolent dictatorships to a leadership group to a more anarchic structure where anyone who is a contributor can contribute to the decision making process. This supports existing literature that finds that in regards to bug fixing there is no distinguishing social structure for OSS projects (Crowston & Howison, 2005).

Participant 17 worked on a project where the leader was not focused on leadership but rather to finish a functionality for her own use. After that aspect of the project was finished she left. Participant said this:

> "So when [the former leader] left, I think the community saw, or the main developers, contributors, saw that it would be good that we keep someone paid to do that. That it was a position that if it was filled with someone, it would help the project be more dynamic. It helps to focus and to get results, because being too organic is kind of bad. People can go in all different directions – then you have no coordination."

While organic structures can be useful, and people with intrinsic motivations can be highly motivated to work on their sections, they also have the downside of being too dynamic and lacking coordination. When projects are too dynamic, this can lead to unclear guidance for contributors. For example, earlier participant 06 noted that because of a lack of guidelines, potential contributors submitted sub-par code, which then had to be rejected. Projects that have unclear submission guidelines demonstrate poor leadership, and inhibit future contributors from wanting to join.

Disrupted leadership can have adverse effects on project functioning (Singer et al., 2004; Wasley, 1992). The demands of the OSS project tied in with the lack of payment and invisible labor led some of our sample to leaving OSS. Most OSS projects have a very low truck factor, meaning that if there are often one or two leaders and if one or both leave then the project cannot maintain itself. Leadership is necessary to help contributors write code in the right directions. We found that leaders sometimes had to stop people from adding functionality that wasn't necessary or when they made modifications that were not going to be possible to maintain. This in and of itself is a form of invisible labor, because stopping people from making contributions (albeit improper ones) is a contribution that will often happen in places like Slack or Discord channels. This social labor underscores the importance of management.

Leadership is also essential for building an active community to recruit and retain future contributors and leaders. This is important for two reasons, first because development of a critical mass of active participants is crucial for sustaining OSS forums (Wasko et al., 2005). Second, recruiting occasional contributors to being active project participants allows them to possibly become leaders themselves. Leaders in our study were not always able to onboard contributors successfully.

**Funding**

Invisible labor additionally underpins the financial sustainability of OSS: projects often need funding to sponsor maintenance work that complements volunteer contributions, but the skills needed to obtain that funding are largely unpaid and under-recognized. The steps to acquire funding are not traditionally taught to people who join OSS projects, and learning them requires more uncompensated work. This labor is invisible and not just unpaid because establishing a committee and legitimacy is not traditionally recognized in OSS projects as a form of labor.

In our study, participants are staying with OSS projects due to intrinsic motivation or because they have personal investment in the success of these projects. This contrasted with prior research which suggests that people participate in hope labor to improve future job prospects, which we did not find. None of the participants in our sample were explicitly interested in using OSS to improve their careers. This does not necessarily mean that this is not a driving motivation for contributors. We primarily interviewed leaders of and long-term contributors to OSS projects for their perspectives on governance, it is very possible that people who are looking to advance their careers are more likely to contribute until they can get a better job and therefore unlikely to be in leadership positions. They may purposefully avoid leadership positions because they are aware of the amount of unpaid invisible labor that will get them less future recognition. In addition, our sample was biased toward projects associated with academia, this meant that contributors may be interested in advancing their career, but only in the sense that the OSS projects can be used to demonstrate their own work.

**Creating Legitimacy to Attract Funding**

Our study contributes to the literature on invisible labor by recognizing another form of invisible labor: the creation of legitimacy to attract funding. Like traditional workplaces, OSS projects need to find their own funding or remain entirely volunteer-run. The creation of legitimacy to obtain that funding is crucial to making an organization seem legitimate, and utilitarian which allows projects to solicit grant funding or corporate sponsorship.

The lack of funding is a well known problem in OSS. Without funding, all contributors are working for free, while with funding, projects can usually fund one or two contributors (a fraction of the total contributors). With funding, projects can usually sponsor a small fraction of their contributors to work on time-consuming or unappealing tasks that volunteer labor rarely satisfies. To solve this problem, some projects try to solicit donations from their users; however, these donations tend to be negligible (Turner, 2021). Other projects try to become nonprofits or attractive to businesses to acquire sponsorship, but this isn't possible for many projects (Stannard, 2020). There is a growing awareness of the need for funding in OSS, and institutions like the National Science Foundation and Chan Zuckerberg Initiative are investing in maintaining OSS ("NSF Invests over $26 Million in Open-Source Projects.", 2023; Coldewey, 2020) most OSS projects are unfunded. Getting funding requires more invisible labor such as knowing how to make your project into a business or writing grants to obtain funding.

**Academic OSS**

Our sample was largely focused on academic OSS or projects with utility for academic research. This was partially because the utility of these projects allowed them to grow to a size that fit our sampling criteria, but their specificity stopped them from becoming behemoths. We also focused on academic OSS because these projects had successfully solicited funding from organizations like the Chan Zuckerberg Initiative meaning that they were on lists of grant awardees, making them easy to find. This also meant that they had created legitimacy around their project, which in turn, meant they often had functioning websites with contributor pages, making it easy to find their contributors.

The focus on academic OSS is useful because this is an area that has very little focus, and broadly, has large applications within academic research. However, this focus on academic OSS does somewhat limit our generalizability. First, because we are using projects that are seeking funding, we are finding projects who need to create legitimacy and submit funding proposals. Projects that are not interested in soliciting funding may have different strategies for long term survivability, or may be unconcerned with longevity. These projects may thus utilize less invisible labor. Second, because we are focusing on academic OSS, projects likely contain graduate students, meaning that the traditional goals of OSS for first time contributors might not be relevant. In our sample, we don't find that people are using OSS to further their careers in software development, but we might see more of this outside of our sample. Projects connected to labs may also not worry about recruiting new members, as people who join the lab may be implicitly required to assist in the OSS project.

**Limitations**

Our participants were in positions of leadership and often had extensive experience. Having been immersed in this ecosystem, our participants are in the perfect position to give us a detailed understanding of the invisible labor they perform and insight into what works for OSS and what does not. The people in our study were primarily funded outside of OSS, and contributed out of a sense of commitment to the ethos of OSS. Many of them were employed from a job not related to OSS and were getting paid a lot more than OSS could ever pay them. Their dedication to OSS means that they may not be representative of a typical user who contributes once to fix a bug or add a feature.

**Acknowledgements**

This project has been made possible in part by grant number 2021-237100 from the Chan Zuckerberg Initiative DAF, an advised fund of Silicon Valley Community Foundation. Additionally, M. Z. Trujillo is supported by the Northeastern University Future Faculty Postdoctoral Fellowship Program.

References

Avelino, G., Constantinou, E., Valente, M. T., & Serebrenik, A. (2019, September). On the abandonment and survival of open source projects: An empirical investigation. In *2019 ACM/IEEE International Symposium on Empirical Software Engineering and Measurement (ESEM)* (pp. 1-12). IEEE.

Coldewey, Devin. "Chan Zuckerberg Initiative Drops $3.8M on 23 Biomedical Open-Source Projects." *TechCrunch*, 4 May 2024, techcrunch.com/2020/05/27/chan-zuckerberg-initiative-drops-3-8m-on-23-biomedical-open-source-projects/.

Bagozzi, R. P., & Dholakia, U. M. (2006). Open source software user communities: A study in participation in Linux user groups. *Management Science, 52*, 1009-1115.

Bonaccorsi, A., & Rossi Lamastra, C. (2003). Altruistic individuals, selfish firms? The structure of motivation in Open Source software. *The Structure of Motivation in Open Source Software*.

Brabham, D. C. (2010). Moving the crowd at Threadless: Motivations for participation in a crowdsourcing application. *Information, Communication & Society, 13*, 1122-1145. doi: 10.1080/13691181003624090

Crowston, K., & Howison, J. (2005). The social structure of free and open source software development. *First Monday, 10*(2). doi: 10.5210/fm.v10i2.1207

Daigle, K., & Github Staff, G. S. (2024, July 30). *Octoverse: The State of Open Source and rise of AI in 2023*. The GitHub Blog. https://github.blog/2023-11-08-the-state-of-open-source-and-ai/

Daniels, A. K. (1987). Invisible work. *Social problems*, 34(5), 403-415.

Demby, G. (2013, December 5). Why Isn'tOpen Source A Gateway For Coders Of Color? *NPR*.

Fang, Y., & Neufeld, D. (2009). Understanding sustained participation in open source software projects. *Journal of Management Information Systems*, 25(4), 9-50.

Finley, A. (2020, May 29). *Women's household labor is essential. why isn't it valued? - The Washington Post*. The Washington Post . https://www.washingtonpost.com/outlook/2020/05/29/womens-household-labor-is-essential-why-isnt-it-valued/

Gadman, L. J. (2013). Open source leadership: Leading and managing community led programmes to support strategies for next generation broadband implementations across Europe. *International Journal of Organizational Analysis,* 21(4), 528-545.

Ghariwala. V. (2023, January 19). *The Significance of Open Source Software in the Digital-First Future Enterprise*. CIO. https://www.cio.com/article/419466/the-significance-of-open-source-software-in-the-digital-first-future-enterprise.html

Gray, M. L., & Suri, S. (2019). *Ghost work: How to stop Silicon Valley from building a new global underclass*. Harper Business.

Halfaker, A., Kittur, A., & Riedl, J. (2011, October). Don't bite the newbies: how reverts affect the quantity and quality of Wikipedia work. In *Proceedings of the 7th international symposium on wikis and open collaboration* (pp. 163-172).

Hergueux, J., & Kessler, S. (2022, April). Follow the leader: Technical and inspirational leadership in open source software. *In Proceedings of the 2022 CHI Conference on Human Factors in Computing Systems* (pp. 1-15).

Hoffmann, M., Nagle, F., & Zhou, Y. (2024). The Value of Open Source Software. *Harvard Business School Strategy Unit Working Paper*, (24-038).

Jensen, C., King, S., & Kuechler, V. (2011, January). Joining free/open source software communities: An analysis of newbies' first interactions on project mailing lists. In *2011 44th Hawaii international conference on system sciences* (pp. 1-10). IEEE.

Kaplan, A. (2022). "Just Let it Pass by and it Will Fall on Some Woman": Invisible Work in the Labor Market. *Gender & Society*, 36(6), 838-868.

Kuehn, K., & Corrigan, T. F. (2013). Hope labor: The role of employment prospects in online social production. *The political economy of communication, 1*(1).

Lakhani, K. R., & Wolf, R. G. (2005). Why hackers do what they do: Understanding motivation and effort in free/open source software projects. In J. Feller, B. Fitzgerald, S. A. Hissam, & K. R. Lakhani (Eds.), *Perspectives on free and open source software* (pp. 322). Cambridge, MA: MIT Press.

Lerner, J., & Tirole, J. (2002). 'Some simple economics of open source. *Journal of Industrial Economics*, *50*, 197–234. doi:10.1111/1467-6451.00174

Li, H., Hecht, B., & Chancellor, S. (2022, May). All that's happening behind the scenes: Putting the spotlight on volunteer moderator labor in reddit. In *Proceedings of the International AAAI Conference on Web and Social Media* (Vol. 16, pp. 584-595).

Majid, A. (2024, August 26). *Editorial note rev. 2.0: Open source as an ideology of participation and inclusiveness*. INVERSE JOURNAL. https://www.inversejournal.com/2024/08/26/editorial-note-rev-2-0-open-source-as-an-ideology-of-participation-and-inclusiveness-by-amjad-majid/

Marmorstein, R. (2011, June). Open source contribution as an effective software engineering class project. In *Proceedings of the 16th annual joint conference on Innovation and technology in computer science education* (pp. 268-272).

Meluso, J., Casari, A., McLaughlin, K., & Trujillo, M. Z. (2024). Invisible Labor in Open Source Software Ecosystems. *arXiv preprint arXiv:2401.06889*.

Microsoft. (2024). *Microsoft's Free and Open Source Software Fund (FOSS Fund)*. Github. https://github.com/microsoft/foss-fund

Nafus, D. (2012). 'Patches don't have gender': What is not open in open source software. *New Media & Society*, 14(4), 669-683.

Neff, G. (2012). *Venture labor: Work and the burden of risk in innovative industries*. MIT press.

"NSF Invests over $26 Million in Open-Source Projects." *NSF*, 25 Oct. 2023, new.nsf.gov/tip/updates/nsf-invests-over-26m-open-source-projects.

Oreg, S., & Nov, O. (2008). Exploring motivations for contributing to open source initiatives: The roles of contribution context and personal values. *Computers in human behavior*, 24(5), 2055-2073.

Rashid, M., Clarke, P. M., & O'Connor, R. V. (2019). A systematic examination of knowledge loss in open source software projects. *International Journal of Information Management*, 46, 104-123.

Riehle, D., Riemer, P., Kolassa, C., & Schmidt, M. (2014, January). Paid vs. volunteer work in open source. In *2014 47th Hawaii International Conference on System Sciences* (pp. 3286-3295). IEEE.

Roberts, S. T. (2019). *Behind the screen*. Yale University Press.

Singer, P., Goodrich, J., & Goldberg, L. (2004). Your library's future: When leaders leave, succession planning can smooth the transitions. *Library Journal,* 129(17), 38-41.

Small, M. L. (2009). How many cases do I need?' On science and the logic of case selection in field-based research. *Ethnography*, *10*(1), 5-38.

Stannard, Aaron. "How to Build Sustainable Open Source Software Projects." *Aaronontheweb Software. Startups. Outer Space* , 30 Jan. 2020, aaronstannard.com/sustainable-open-source-software/.

Steinmacher, I., Silva, M. A. G., Gerosa, M. A., & Redmiles, D. F. (2015). A systematic literature review on the barriers faced by newcomers to open source software projects. *Information and Software Technology*, *59*, 67-85.

Steinmacher, I., Wiese, I., Chaves, A. P., & Gerosa, M. A. (2013, May). Why do newcomers abandon open source software projects?. In *2013 6th International Workshop on Cooperative and Human Aspects of Software Engineering (CHASE)* (pp. 25-32). IEEE.

Turner, James. "Open Source Has a Funding Problem." *Stack Overflow*, 7 Jan. 2021, stackoverflow.blog/2021/01/07/open-source-has-a-funding-problem/.

Wasko, M. M. & Faraj, S. F. (2005). Why should I share? Examining social capital and knowledge contribution in electronic networks of practice. *MIS Quarterly, 29*, 35-57.

Wasley, P. A. (1992). When Leaders Leave. *Educational Leadership*, 50(3), 64-67.

Winter, R., Salter, A., & Stanfill, M. (2021). Communities of making: Exploring parallels between fandom and open source. *First Monday.*

Wodecki, B. (2024, February 6). *GitHub: Open source projects risk dying without more funding*. AI Business. https://aibusiness.com/ml/github-open-source-projects-risk-dying-without-more-funding-

Zuegel, Devon. "Let's Talk about Open Source Sustainability." *The GitHub Blog*, 10 Jan. 2022, github.blog/2019-01-17-lets-talk-about-open-source-sustainability/.